\documentclass[a4paper,11pt]{article}
\pdfoutput=1 

\usepackage{jcappub} 

\usepackage[T1]{fontenc} 

\usepackage{float}

\usepackage{natbib}
\usepackage{hyperref}

\usepackage{subfiles}
\usepackage{aas_macros}
\usepackage{amsmath}
\usepackage{amssymb}
\usepackage{tabularx}
\usepackage[section]{placeins}
\usepackage{graphicx}
\usepackage{booktabs}
\usepackage[normalem]{ulem}
\usepackage{aas_macros}
\useunder{\uline}{\ul}{}

\newcommand{\unit}[1]{\ensuremath{\, \mathrm{#1}}}
\newcommand{\vbar}{\bar{v}}
\newcommand{\vesc}{v_\text{esc}}

\newcommand{\be}{\begin{equation}}
\newcommand{\ee}{\end{equation}}

\newcommand{\mx}{m_\chi}
\newcommand{\rhox}{\rho_\chi}

\newcommand{\Robj}{R}

\newcommand{\Kn}{\text{Kn}}

\title{Constraining Asymmetric DM Properties by Black Hole Formation in Neutron Stars and Population III Stars}

\author{Jared Diks, Cosmin Ilie}

\affiliation{Department of Physics and Astronomy, Colgate University,\\13 Oak Dr., Hamilton, NY 13346,  U.S.A.}

\emailAdd{jdiks@colgate.edu}
\emailAdd{cilie@colgate.edu}

\begin{document}
\abstract{In this work we explore the potential for Neutron Stars (NSs) at the Galactic center and Population~III stars to constrain bosonic Asymmetric Dark Matter (ADM). We demonstrate that for NSs in an environment of sufficiently high DM density ($\rhox\gtrsim10^{9}\unit{GeV/cm^3}$), the effects of both multiscatter capture and DM evaporation cannot be neglected. Conversely, for Pop~III stars,  we find they are excellent at probing low-mass ADM. For instance, the most easily observable Population III stars could be highly effective at constraining high-$\sigma$ low-$\mx$ DM, maintaining efficacy below $\mx=10^{-15}\unit{GeV}$ (assuming a Bose Einstein Condensate(BEC) forms) thanks to their far lower value of $\mx$ at which capture saturates to the geometric limit. Finally, we derive closed-form approximations for the evaporation rate of DM from arbitrary polytropic objects and from DM particles in a BEC state.}

\maketitle

\flushbottom

\section{Introduction}
Dark Matter is one of the most significant open problems in modern physics. Since the Coma Cluster was observed by Fritz Zwicky in the 1930's \cite{1933AcHPh...6..110Z}, evidence for an invisible form of matter distributed throughout the universe has continued to accumulate. Proposed solutions to this include two main hypotheses: either gravity works differently than Newton and Einstein proposed, or Dark Matter is a particle that interacts with the standard model primarily through gravity, and rarely (if at all) through other means. In recent years, the former (Modified Newtonian Dynamics, or MOND) hypothesis has been largely ruled out \cite{10.1093/mnras/stad3393}, leaving the particle hypothesis of Dark Matter as the most probable explanation.

In light of the scarcity of the clues about DM's properties, many particle models have been proposed. The most popular of these are the WIMP, SIMP, and Co-SIMP, all of which are "symmetric" models, meaning that the DM particle is its own antiparticle. However, of particular interest to this work, several "asymmetric" (ADM) models exist. While more standard DM models such as the WIMP require a "cosmic coincidence" to explain the similar (to order of magnitude) densities of DM and Baryons in the universe, the relic density of DM is set by the baryon-antibaryon asymmetry (as opposed to thermal freeze-out which defines WIMP density), giving it the name "asymmetric DM" \cite{Kaplan_2009}. Since there is no clear answer as to whether our universe contains symmetric and/or asymmetric DM, our analysis assumes that asymmetric DM exists, and predicts detectable consequences so we can narrow down the possibilities of what the particle's properties may be, should it exist.

Regardless of the model considered, physicists take both direct and indirect approaches to detection. The direct approach involves building detectors on Earth such as the XENON \cite{Aprile_2017}, DM-Ice \cite{pettus2015dmicecurrentstatusfuture}, and CRESST \cite{kluck2017searchlowmassdarkmatter}. However, this method has several limitations: building and maintaining these experiments is expensive, and they have a systematic limitation known as the "neutrino floor" or "neutrino fog", a sensitivity limit beyond which any DM interaction would be drowned out by neutrino signals. In other words, DM and Neutrinos are detected through similar Weak-Force interactions, so any direct detector that could theoretically probe for sufficiently low-mass DM would detect so many neutrinos that no DM interaction would be distinct enough to be accurately measured.

To bypass these limitations, several indirect detection methods have been devised, the majority of which utilize observations of celestial bodies to draw conclusions about the DM that may have been "captured" by them over time. With symmetric DM, this generally involves looking for signatures of excess heating resulting from DM annihilations, applications of which range from Exoplanets \cite{Ilie:2023lbi, Leane:2020wob}, to Population III stars \cite{Ilie:2020PopIII, Freese:2008cap}, to White Dwarfs \cite{Horowitz:2020axx, Dasgupta:2020dik} and Neutron Stars\cite{Bramante:2017, Ilie:2020Comment, Nguyen:2022NS, Baryakhtar:2017, Bell:2020,Bell:2018pkk}. As this method is inapplicable to ADM, detection of ADM requires taking advantage of the main consequence of its stability: captured ADM will never annihilate within a star, so in the absence of any other escape method, ADM will pile up within celestial bodies without bound. In optimal conditions, this accumulation could grow so dense that it collapses into a black hole, consuming the progenitor star along with it. As such, by deriving when this occurs based on stellar parameters, we can place bounds upon the scattering cross section and mass of the ADM particle.

As it turns out, formation of a black hole (BH) through ADM collapse is most useful in a single scenario: scalar ADM in a neutron star (NS). Scalar DM is special, because due to its spin of 0, it has no exclusion principle, meaning that the only thing stopping any self-gravitating mass of ADM from forming a BH is the pressure due to zero-point energy, which is orders of magnitude smaller than the Pauli pressure that fermions experience. Neutron stars in particular form the perfect environment for ADM collapse for a few reasons: their small size increases the gravitational attraction between DM particles, their incredible density results in highly efficient DM capture, and their high escape velocity and relatively low central temperatures limits the effects of DM evaporation, the opposite effect of capture. 

Though they are not nearly as effective as a NS in identical conditions, we find that early-universe Population III stars can also function as ADM detectors: despite their much lower density and larger size, the ambient DM density in the early universe was far higher than it currently is, allowing Population III stars (in a DM halo with density $\rhox=10^{13}\unit{GeV/cm^3}$) to detect ADM approximately as effectively as a NS near the galactic center ($\rhox=10^{6}\unit{GeV/cm^3}$). In addition to consideration of new objects, we make several improvements upon the methodology pioneered by \cite{McDermott_2012} by widening the range of considered values of the ambient DM density to the maximum theorized at the galactic core, including the full effects of evaporation and multiscatter DM capture, and applying a new capture suppression effect\cite{lu2024} when $\mx$ is large such that the energy of the collision is high enough that the impacted nucleon can no longer be considered a point particle.

\section{Scalar Dark Matter Processes in Stars}
There are two main DM processes of relevance to asymmetric DM: capture and evaporation. Capture is the process by which DM particles collide with baryonic particles within the star, occasionally losing enough momentum to fall below the escape velocity of the star, becoming gravitationally bound to it~\cite{1988ApJ...328..919G,Bramante:2017,Ilie:2020Comment}. The reverse of this process is known as evaporation, in which DM particles which are already gravitationally bound to the star are sufficiently accelerated through collisions to escape the gravitational well of the star~\cite{1987ApJ...321..560G}. Considering these two processes, we are able to model the total number of DM with the following differential equation: 

\begin{equation}
    \frac{dN_X}{dt} = C - E N_X.
    \label{eq:dNx_DiffEq}
\end{equation}

\noindent where $N_X$ is the total number of DM particles gravitationally bound to the star, C is the capture rate, and E is the per-particle evaporation rate. As we will show, in most of our considered parameter space, evaporation is irrelevant due to the high escape velocity of Neutron Stars. However, in areas of high DM density, we cannot ignore the effects of evaporation, and will use the full numerical solution to Eq.~\ref{eq:dNx_DiffEq}. Whenever the per particle evaporation rate ($E$) is independent of the number of DM particles ($N_X$) one can find the following analytic solution for $N_X(t)$:

\begin{equation}
    N_X = \frac{C}{E}(1-e^{-Et})
    \label{eq_Nx_solved}
\end{equation}

Following the example of \cite{McDermott_2012} we will consider a typical old NS with $t=10^{10}$ years, $M=1.44M_\odot$, $\rho_B = 1.4\times 10^{15} \unit{g/cm^3}$ to stay consistent with \cite{McDermott_2012}.

\subsection{Multiscatter Capture}
Though a DM particle may collide once with a neutron in the NS an lose enough momentum to be captured, its velocity may not be decreased enough to fall below the escape velocity. In this case, the particle may escape the star, but may instead collide with another neutron, losing even more momentum. This process can proceed indefinitely, until the particle is either captured or escapes. As such, the total capture rate can be calculated as \cite{Bramante:2017}:

\begin{equation}
\resizebox{.92\hsize}{!}{$C = \sum^\infty_{N=1}C_N = \sum^\infty_{N=1} \underbrace{\xi F(\delta p^2)}_\textrm{supp. factors} \underbrace{\pi \Robj^2}_\textrm{capture area}\times \,\underbrace{n_X \int_0^{\infty} \dfrac{f(u)du}{u}\,(u^2+\vesc^2)}_\textrm{DM flux}\times \, \underbrace{p_{ N}(\tau)}_\textrm{prob. for $N$ collisions}\times \, \underbrace{g_{ N}(u)}_\textrm{prob. of capture},$}
\label{eq:fullCap}
\end{equation}

\noindent where $C_N$ represents the capture rate of DM particles that collide exactly $N$ times. In the case of the neutron star, neutron degeneracy effects act to suppress capture for $\mx \lesssim 1 \unit{GeV}$, accounted for by the factor $\xi_{s}$, which is parameterized as:

\begin{equation}
    \xi_{s} = \text{Min}\left[\frac{\delta p}{p_F} ,1 \right],
\end{equation}

\noindent where $\delta p \approx \sqrt{2}m_r v_{esc}$ represents the expected momentum transfer between a DM particle and a degenerate neutron with $m_r$ as the reduced mass of the DM particle, and $p_F\simeq(3 \pi^2 \rho_B/m_n)\simeq 0.575 \unit{GeV}$ is the Fermi momentum for a NS of parameters defined in section 1, as noted by \cite{McDermott_2012}. As a result, capture below $\sim1\unit{GeV}$ will be suppressed by a factor of $\sim\mx \vesc / p_F$.

Whenever $\mx$ is large the energy transferred to the neutron in capturing collisions is also large enough that we can no longer consider the interaction as a point particle~\cite{lu2024}. To account for this, we include the form factor:

\begin{equation}
    F(\delta p^2) = \frac{\Lambda^4}{(\delta p^2 + \Lambda^2)^2}
    \label{eq:form_factor}
\end{equation}

\noindent where we follow \cite{lu2024} in choosing $\Lambda \approx 0.25\unit GeV$ based on the assumption that the typical nucleon is approximately $0.8 \unit{fm}$ in radius. As we will show, the form factor acts to suppress the boundary of BH formation by approximately two orders of magnitude when $\mx>m_n$.

\begin{figure}
        \centering
        \includegraphics[width=0.55\linewidth]{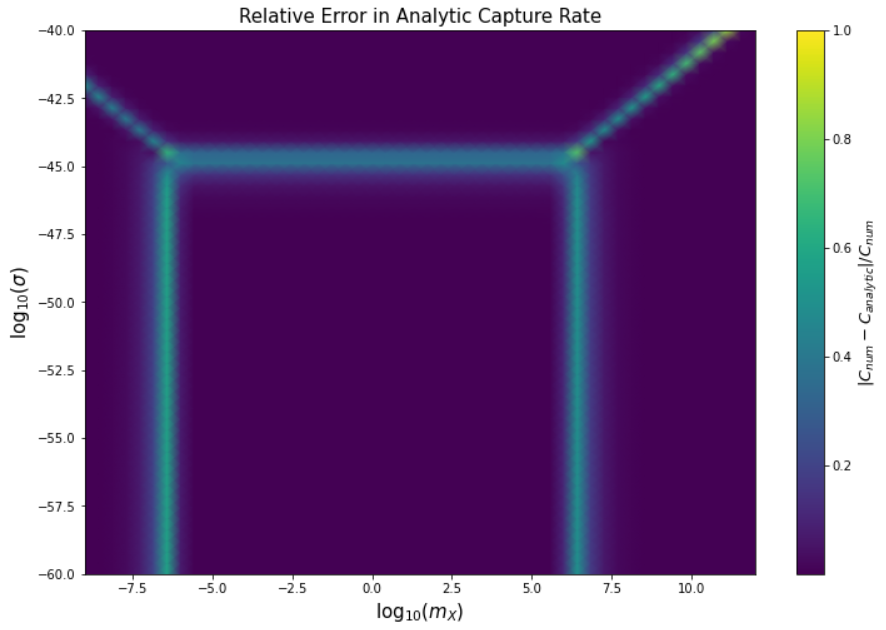}
        \includegraphics[width=0.4\linewidth]{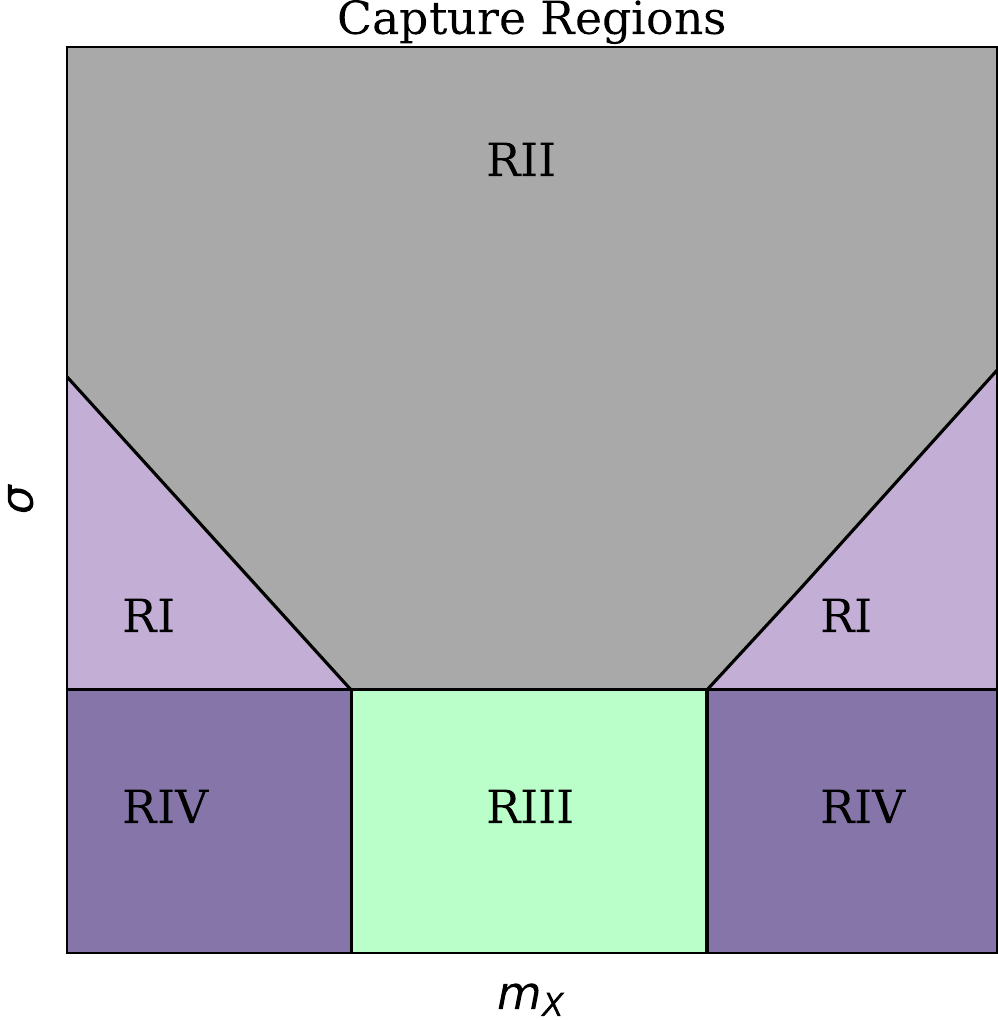}
        \caption{By plotting the relative error between the full capture rate and the regional approximations, the locations of each of these regions for a NS of relevant parameters (and $\rhox = 10^6 \unit{GeV/cm^3}$) becomes clear. The reader may notice that while most of the lines in the right schematic are visible in the left panel, there does not appear to be a line between regions I and IV. This is no coincidence, as remarkably, despite Region IV being within the single-scatter regime and Region I multiscatter, their capture approximations in Equations \ref{eq:CtotRI} and \ref{eq:CtotRIV} are identical. Further, there is some offset between the location of the lines in the right and left panel, which we include to indicate that the location of the boundaries between regions is dependent on the properties of the star.}
        \label{fig:ctot_rel}
\end{figure}

In \cite{Ilie:2023lbi} and \cite{2024IlieCapture}, we showed that depending on the "region" of $\sigma-\mx$ parameter space we consider, the full multiscatter capture rate from Eq.~\ref{eq:fullCap} (excluding the suppression factors, which are NS-specific) reduces to one of four approximate forms. Considering these suppression factors together, since both are constant with respect to $N$, they can be factored out of the sum, and will enter the capture approximations in $\sigma - \mx$ parameter space as defined in \cite{Ilie:2023lbi} as extra prefactors in each region, becoming:

\begin{equation}
    C = 2 \xi_{s} F(\delta p^2) A k\tau \vesc^2, \quad\quad \text{(Region I)},
    \label{eq:CtotRI}
\end{equation}
\begin{equation}
    C = A \xi_{s} F(\delta p^2) \left(2 \vbar^2 + 3\vesc^2\right), \quad \text{(Region II)},
    \label{eq:CtotRII}
\end{equation}
\begin{equation}
    C = 2 \xi_{s} F(\delta p^2) A \tau \vesc^2, \quad\quad \text{(Region III)},
    \label{eq:CtotRIII}
\end{equation}
\begin{equation}
    C = 2 \xi_{s} F(\delta p^2) A k\tau \vesc^2, \quad\quad \text{(Region IV)}
    \label{eq:CtotRIV}
\end{equation}

\noindent where $A\equiv \frac{1}{3}\pi \Robj^2 \sqrt{\frac{6}{\pi}} \frac{n_X}{\vbar}$, $k \equiv \frac{3 \vesc^2}{2 \vbar^2} \langle z\rangle \frac{4 m m_X}{(m + m_X)^2}$ is a dimensionless intermediate variable, $\tau \equiv 2 \Robj \sigma n_B$ is the optical depth of the star in the radial direction, and $\vesc$ is the escape velocity from the surface, and $n_B$ is the average number density of the stellar baryons. A visualization of the location of these regions within the parameter space we consider is given in Figure \ref{fig:ctot_rel}.

\subsection{Evaporation}
We can determine the relevance of evaporation in the considered regions of parameter space by consulting \cite{Garani_2022}, who find that the evaporation mass for a NS of similar parameters to those considered in this paper is $1.4\unit{KeV}$ (note that the NS chosen in \cite{Garani_2022} has a lower mass and identical temperature, so this is an overestimate when applied to our NS with $M=1.44M_\odot$). If $\rhox$ is sufficiently high, the boundary reaches below $\mx=1.4\unit{KeV}$, so we must consider evaporation. The full per-particle evaporation rate is~\citep{Garani_2022}:

\begin{equation}
    E = \frac{1}{N_X}\int_0^{\Robj} s_\text{evap}(r) n_X(r)~4\pi r^2~dr \int_0^{v_e(r)} f_X(\boldsymbol{w}, r)~4\pi w^2~dw \int_{v_e(r)}^\infty \mathcal{R}^+ (w\rightarrow v)~dv,
    \label{eq:Evaporation_IntegralForm}
\end{equation}

\noindent where $n_X(r)$ is the number density of DM particles at a radius $r<R$, $v_e(r)$ is the escape velocity of a particle at a radius $r<R$, $\mathcal{R}^+ (w\rightarrow v)$ is the rate at which particles are upscattered from velocity $w$ to $v$, and $f_X(\boldsymbol{w}, r)$ is the velocity distribution of the particles at that radius. For particles not in the ground state of a BEC, we assume a Maxwell-Boltzmann distribution truncated at the escape velocity of that radius. 
When a non-negligible fraction of the DM in the star is in the BEC state, we must treat those particles separately, applying an independently calculated evaporation rate $E_{BEC}$ to the particles in the BEC. For this purpose we developed an analytic formalism of calculation of $E_{BEC}$, presented in detail in Appendix B. The total evaporation rate then becomes:

\begin{equation}\label{eq:Evap_sum}
    E = E_{BEC} \frac{N_\chi^0}{N_\chi} + E_{MB}\frac{N_\chi-N_\chi^0}{N_\chi}
\end{equation}

\noindent where $E_{MB}$ is the evaporation rate from the standard Maxwell-Boltzmann distribution throughout the star (see details below and in Appendix A), and $N_\chi^0$ is the number of particles in the BEC, the calculation of which will be explained in the next section. In effect, this equation adds together the two evaporation rates weighted by the fraction of particles in their respective distributions to obtain an evaporation rate for the entire star. Notice that with this separation, $E$ is no longer independent of $N_\chi$. In this case, Equation \ref{eq:dNx_DiffEq} has no closed form solution, so we solve it numerically. However, for both types of objects considered in this paper (NS and Pop~III stars) whenever evaporation becomes relevant one of the two terms dominates. For instance, we find that for NSs, in view of their cooler core temperature, even when they form a BEC, the DM evaporation is mostly due to the small number of particles left in the non ground state. Conversely, for Pop~III stars, wich have a much hotter central temperature, once a BEC forms and for $m_X$ sufficiently low, the evaporation is dominated by the $E_{BEC}$ term.

Further assuming a velocity-independent $\sigma$, the rate of up-scattering is parameterized the following way \cite{Garani:2017}:

\begin{equation}
    \mathcal{R}^+ (w\rightarrow v) = \mathcal{R}^+ (w,~v,~m,~m_X,~T(r),~n(r),~\sigma),
\end{equation}

\noindent where $n(r)$ is the radial density profile of neutrons within the NS. Throughout this paper, we make the simplifying assumption that the entire mass of the NS is neutrons distributed according to an $n=1.5$ polytrope, which effectively models degenerate objects. To convince yourself of this, note that $n=1.5$ is the index which reproduces $P\sim\rho^\frac{5}{3}$ in the equation of state for a degenerate neutron gas.

Finally, $s_\text{evap}(r)$ accounts for evaporation suppression, a process which in Neutron Stars can significantly limit the evaporation rate. When a captured DM particle collides with a neutron and is accelerated above the escape velocity of the star at that radius, it is possible that the particle impacts another neutron before escaping, again falling below the escape velocity.  The suppression factor $s_\text{evap}(r)$ accounts for this, and is given by \cite{Garani:2017}:

\begin{equation}
    s_\text{evap}(r) = e^{-\tau(r)} \underbrace{\frac{7}{10} \frac{1 - e^{-\frac{10}{7}\tau(r) }}{\tau(r)}}_{\text{Angular Factor}} \underbrace{{}_0 F_1\left(; 1 + \frac{2}{3} \hat{\phi}(r); \tau(r)\right)}_{\text{Multiscatter Factor}},
    \label{eq:SuppressionFactor}
\end{equation}

\noindent where the angular factor accounts for the fact that DM particles move in non-radial orbits, and the multiscatter factor represents the probability that a DM particle scatters multiple times, but still manages to escape. In the multiscatter factor, $_0F_1(;b;z)$ is the confluent hypergeometric limit function, and $\hat\phi(r) \equiv \frac{\mx v_e^2(r)}{2 T(r)}$ is the local dimensionless escape energy. Throughout, $\tau(r)$ represents the local optical depth in the radial direction.

Of course, it is necessary to calculate $n_X(r)$, which depends on the mass distribution of the star, its temperature profile, $\mx$, and $\sigma$, and can have very different characterizations based on these factors. In Appendix A, we discuss the possible distributions in detail, but we will provide the main conclusions here: once DM is captured, it achieves one of two distributions depending on how frequently the DM interacts with the baryons within the object. If interactions are frequent, the distribution is called "Local Thermal Equilibrium (LTE)", since the DM particles fully thermalize with the baryons at their local radius from the stellar core. Conversely, if interactions are rare, the DM achieves an "Isothermal" distribution, in which the temperature of the DM at any radius from the core is some fraction of the stellar core temperature (ranging from 1 near the core to $\sim$0.6 near the surface). For our considered parameter space in neutron stars, the distribution is fully isothermal, but a non-negligible region of high-$\sigma$ parameter space enters in other objects enters the LTE regime. So, since both regimes must be considered, the two distributions as derived in Appendix A are:

\begin{equation}
    n_{X}^{\text{ISO}} = n_X(0) e^{-\mx\Phi(r)/T_X}, \text{ Isothermal}
    \label{eq:nxIsoNotAppendix}
\end{equation}

\begin{equation}
    n_X^{LTE}(r)= n_X(0) \theta^{p}(\xi), \quad p = \frac{\xi_1}{\alpha_n}\frac{E_e}{T_c} + \frac{3}{2} - \alpha_0, \text{ LTE,}
    \label{eq:nxLTE_maintext}
\end{equation}

\noindent which we follow \cite{Garani:2017} in stitching together according to the Knudsen Number as:

\begin{equation}
    n_X(\xi) = f(\text{Kn}) n_{X}^{\text{LTE}}(\xi) + \left[1-f(\text{Kn})\right]n_{X}^{\text{ISO}}(\xi),
    \label{eq:nx_r_Transition_inText}
\end{equation}

\noindent where $f(\text{Kn})$ is a transition function defined in \ref{eq:KnudsenTransitionFunction}. As it turns out, plugging \ref{eq:nx_r_Transition_inText} into \ref{eq:Evaporation_IntegralForm} simplifies to a similar stitching expression for the evaporation rates from either regime, allowing us to use approximations \ref{eq:Evap_Iso} and \ref{eq:Evap_LTE} to efficiently calculate the total evaporation rate from particles not in the BEC:

\begin{equation}
    E_{MB} = f(\text{Kn}) E_{\text{LTE}} + \left[1-f(\text{Kn})\right] E_{\text{ISO}},
    \label{eq:Evaporation_interpolated_intext}
\end{equation}

\noindent With this expression in hand, we are able to directly calculate $N_X$ with \ref{eq_Nx_solved} for application to black hole formation.

\section{Derivation of Bounds on $\mx$ and $\sigma$ by NS Destruction}\label{sec:DerivBoundsNS}

In order for a BH to form from asymmetric bosonic DM within a NS, several conditions must be met: the DM must thermalize with the neutrons of the star (losing enough energy to sink to the core), the DM must be numerous enough to self-gravitate, and numerous enough to surpass the Chandrasekhar limit and collapse. In the case that we do not consider the formation of a BEC, \cite{McDermott_2012} shows that the governing condition is self-gravitation, as the number of DM required to self-gravitate is greater than the number required to collapse to a BH ($N_{self}>N^{boson}_{Cha}$). So, the condition for BH formation is $N_\chi > N_{self}$. This quantity is given by\cite{McDermott_2012}:

\begin{equation}
    N_{self} = \frac{4 \pi \mx}{3 r_X^3 \rho_c}, 
\end{equation}
\begin{equation}
    \text{where } r_X=\sqrt{\frac{9T}{4\pi G \rho_c \mx}}
    \label{eq:rxIso}
\end{equation}

Note that we have decided to use the notation originally proposed by \cite{1985Spergel} to call attention to the fact that this expression for $r_X$ is only necessarily valid if we assume the NS to have uniform density and temperature, something not noted in \cite{McDermott_2012}. Since this assumption is valid to order of magnitude, we will continue with caution. Further complicating the use of this approximation of $r_X$ is the inherent assumption that the DM is isothermally distributed. However, despite the expected difficulties in using \ref{eq:rxIso} throughout the whole parameter space, we were able to verify numerically that the difference in the fraction of DM distributed below $r_X$ between the Isothermal and LTE distributions is always within 1\%, so we conclude that Eq.~\ref{eq:rxIso} is an excellent approximation throughout the considered parameter space.

Following \cite{McDermott_2012}, whenever a BEC can form, it is possible for the BEC to self-gravitate before overcoming the Chandrasekhar limit. Since this turns out to be true for all of our considered parameter space, and the only particles self-gravitating are those within the BEC, the condition for BH formation becomes $N^0_\chi > N^{boson}_{Cha}$, where $N^0_\chi$ is the number of particles in the condensed ground state. If the BEC forms (which occurs at $T_{DM}<T_{crit}$), we can calculate the number of particles in the ground state as:

\begin{equation}
    N^0_\chi = N_X \left[1 - \left(\frac{T_{DM}}{T_{crit}}\right)^\frac{3}{2}\right],
    \label{eq:Nx0}
\end{equation}

\noindent where $T_{DM}$ is the temperature of the DM, and $T_{crit}=\frac{2\pi}{\mx}\left[\frac{n_X}{\zeta(3/2)}\right]^{\frac{3}{2}}$ is the critical temperature to form a BEC, with $\zeta$ representing the Riemann Zeta function. To calculate $T_{DM}$ in the isothermal case for the NS, we follow \cite{McDermott_2012} in equilibrating it to $T_c$. In the Population III case, we follow equation B12 in \cite{Ilie:2020PopIII} to find that $T_{DM}\approx T_c$ for $\mx\gtrsim1\unit{GeV}$, and $T_{DM}\approx0.6T_c$ for $\mx\lesssim10^{-2}\unit{GeV}$. In the LTE case, since the DM distribution is highly cored in relevant parameter space, $T_{DM}\approx T_c$. This is an important distinction that has been omitted in previous literature. Next, as derived in \cite{McDermott_2012}, the Chandrasekhar limit in the case of Bosonic DM does not depend on radius, so we can parameterize it as a number of DM particles, corresponding to the number of DM particles a star must capture for the DM to collapse into a black hole:

\begin{equation}
    N^{boson}_{Cha} \approx \left(\frac{M_{pl}}{\mx}\right) \approx 1.5\times 10^{34} \left(\frac{100\unit{GeV}}{\mx}\right)^2
\end{equation}

Altogether, these equations are enough to calculate the upper boundary line in $\sigma-\mx$ parameter space with the full effects of multiscatter capture and evaporation included, but there is one assumption we must check before we continue. In order to have a predictable DM distribution throughout the star such that we can derive the location of the bound, we assume that the DM particles have thermalized. If this is not the case, we can no longer guarantee that a BH would form, so we must lift our bounds. The assumption is valid for much of the parameter space we consider, but is dependent on the age and temperature of the star, as well as $\sigma$ and $\mx$, with a closed-form approximation for the thermalization timescale again provided by \cite{McDermott_2012}:

\begin{equation}
    t_{th} \approx \frac{\mx^2 m_n p_F}{4\sqrt{2} n_c \sigma m_r^3}\frac{1}{E_{th}}
\end{equation}

\noindent where $E_{th}\equiv 3T_DM/2$ is the expected DM energy after thermalization. So, if the age of the star $t<t_{th}$, then the bounds must be lifted. Fortunately, for NS, the limitation imposed by the thermalization requirement is weaker at low-$\mx$ than high, where direct detection experiments are more equipped to detect particles. As we will see, thermalization is not a major consideration for old NS, but can be considerable for younger objects.

It is possible depending on the mass of the black hole that forms, the BH will evaporate before consuming the host star. If a BEC does not form, \cite{McDermott_2012} estimates a critical BH mass $M_{crit}^{BH}\approx1.2\times10^{37} \unit{GeV}$ above which the BH accretes more quickly than it evaporates. Since our boundary condition for BH formation with no BEC is $N_\chi>N_{self}$, then if $N_{self}\mx>M_{crit}^{BH}$, the BH will continue to accrete stellar material until it is consumed entirely. The region in which the BH evaporates before consuming the star is the diamond-hatched region in the left panel of Fig. \ref{fig:bounds_M1000}. If, instead, a BEC forms, \cite{McDermott_2012} finds that BEC formation allows the newly formed BH to accrete DM much more efficiently, pushing the boundary at which the BH evaporates to much higher $\mx$. However, the Hawking radiation may interfere with the formation of the BEC by heating it, and the magnitude of this effect is dependent on the model of ADM being considered. So, just as \cite{McDermott_2012}, we have chosen to mark the region above $\mx=13 \unit{GeV}$ at which the constraints become model-dependent. For a more in-depth analysis of how they arrived at this, check section 5-b of their paper.

\subsection{Analytic Bound Locations}
Depending on the capture region the bound is within, whether capture is suppressed, and whether or not a BEC forms, the form of the boundary will take one of the following forms. To derive these, we simply solve either $N^0_\chi > N^{boson}_{Cha}$ or $N_\chi > N_{self}$ depending on whether we are considering a BEC or not, respectively. In doing so, we require values for $C$, which we obtain by choosing the relevant capture approximation  for the desired region of parameter space from Eqs. \ref{eq:CtotRI}-\ref{eq:CtotRIV}. First, if no BEC forms:

\begin{equation}
    \sigma_\text{min} = D \frac{p_F}{\sqrt{2}\vesc}\mx^{-\frac{5}{2}},  \quad \text{(Region III, $\xi_{s}=\frac{\delta p}{p_F}$)}, 
\end{equation}
\begin{equation}
     \sigma_\text{min} = D \mx^{-\frac{3}{2}},  \quad \text{(Region III, $\xi_{s}=1$)}, 
\end{equation}
\begin{equation}
     \sigma_\text{min} = D \frac{1}{\alpha\langle z \rangle}\frac{1}{m_n}\mx^{-\frac{1}{2}},  \quad \text{(Region IV, $\xi_{s}=1$)}, 
\end{equation}

\noindent where $D \equiv \frac{9}{2\sqrt{6}} \left( \frac{T}{G}\right)^{\frac{3}{2}}\frac{\vbar}{\vesc^2} \frac{1}{\rhox F(\delta p^2) \sqrt{\rho_c}N_B t}$, and $\sigma_{min}$ represents the minimum value of sigma at which a BH forms. Since no realistic DM density would allow the boundary to reach the low-$\mx$ Region IV or Region I if a BEC does not form, these expressions span the full parameter space when capture is not saturated. Conversely, if a BEC is allowed to form, the boundaries become:

\begin{equation}
    \sigma_{\text{min}}=D_\text{BEC}c_1\mx,  \quad \text{(Region III, $\xi_{s}=1$)}, 
\end{equation}
\begin{equation}
    \sigma_{\text{min}}=D_\text{BEC} \frac{p_F}{\sqrt{2}\vesc}c_2\mx^{-2},  \quad \text{(Region III, suppressed)}, 
\end{equation}
\begin{equation}
    \sigma_{\text{min}}=D_\text{BEC} \frac{1}{6\sqrt{2}}\frac{p_F m_n}{\langle z \rangle}c_2 \mx^{-3} ,  \quad \text{(Regions IV and I, suppressed)}
\end{equation}
\noindent where $D_\text{BEC}\equiv \sqrt{\frac{\pi}{6}} \frac{\vbar}{\vesc^2}\frac{1}{\rhox F(\delta p^2) N_B t}$, $c_1=\frac{4\pi\zeta\left(\frac{3}{2}\right)}{3}\left( \frac{9T_{DM}^2}{8\pi^2G\rho_c} \right)^{\frac{3}{2}}$ (where $\zeta$ is the Reimann Zeta function), and $c_2=1.5\times10^{38} \unit{GeV^2}$, which stems from the definition of $N^{boson}_{cha}$.

\section{Bounds from Neutron Stars}

The method for placing bounds on $\sigma$ and $\mx$ is as simple as observing a single star, as the mere fact that the star has not collapsed into a black hole by the mechanisms in the previous sections discounts regions of parameter space. In this section, we will assume that we have observed a NS with $T_c=10^5$K, age $10^{10}$ years, $M=1.44M_\odot$, and $R=10.6$ km, and determine which areas of parameter space a BH would have consumed the star in. Thus, the ADM particle must have values of $\mx$ and $\sigma$ somewhere outside that region. Of course, if the BH forms but evaporates before it consumes the star, the bounds must be lifted (diamond-hatched regions in left panels of subsequent figures. Equation 32 in Ref.~\cite{McDermott_2012} controls the interplay between BH accretion and evaporation, accompanied by a discussion of its effect on the bound-setting process), and likewise we must lift the bounds if Hawking radiation may interfere with DM thermalization (square hatched regions in right panel of subsequent figures).

Since the depth of bounds we can place on $\mx$ and $\sigma$ is largely determined by $\rhox$, we will consider several different possible values of $\rhox$ corresponding to different locations within the Milky Way. A typical estimate for the DM density near the galactic center is $\rhox = 10^6\unit{GeV/cm^3}$ within the inner parsec\cite{Navarro_1996}, which is the highest considered by \cite{McDermott_2012}. However, depending upon the effects of adiabadic contraction, $\rhox$ could go as high as $10^{13}\unit{GeV/cm^3}$ \cite{immortalStars}. Regardless, we will consider this value of $\rhox$ as a possibility, as its effects on our analysis are particularly interesting.

\begin{figure}[H]
    \centering
    \includegraphics[width=1\linewidth]{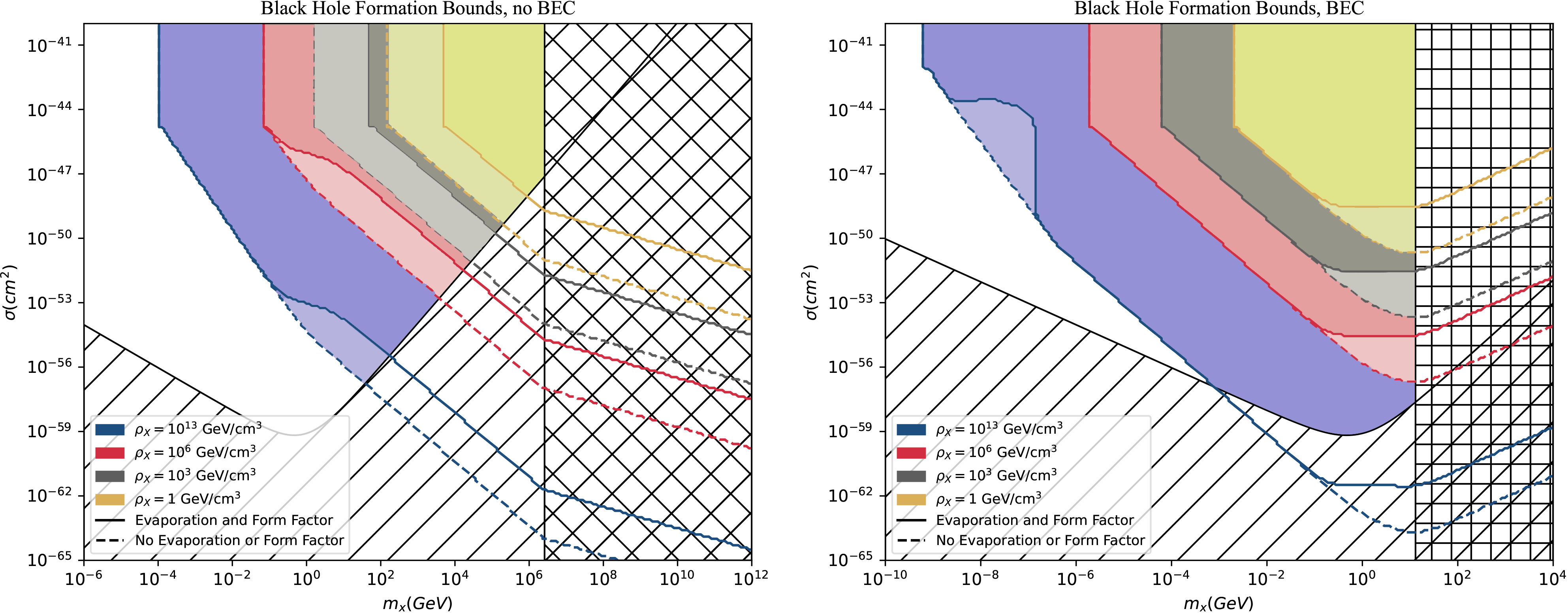}
    \caption{Upper bounds placed on $\sigma$ and $\mx$ by observation of a NS with $t=10^{10}$ years, $M=1.44M_\odot$ at four possible values of $\rhox$. In the diagonally shaded regions, the DM has not had sufficient time to thermalize with the baryons within the star, so the bounds are lifted\cite{McDermott_2012}. Similarly, in the diamond-hatched region, the BH may evaporate before the star collapses, and in the square-hatched region, hawking radiation may disrupt thermalization of the DM. In either case, the bounds are lifted. The dotted colored lines represent the bounds we would get if we did not consider evaporation or the capture suppression form factor (Eq.~\ref{eq:form_factor}) and exist to offer a visual comparison to the bounds derived by \cite{McDermott_2012}. The solid colored lines represent our most accurate bounds including both evaporation and Equation \ref{eq:form_factor}. Any dark shaded area is excluded by observation of a $t=10^{10}$ years old NS of $M=1.44M_\odot$ with $\rho_B = 1.4\times 10^{15} \unit{g/cm^3}$} placed in an area where $\rhox$ is equal to the corresponding value in the legend. Lightly shaded regions would be excluded by an analysis not considering evaporation or the DM capture suppression form factor of (Eq.~\ref{eq:form_factor}).
    \label{fig:bounds_NS}
\end{figure}

In Figure \ref{fig:bounds_NS}, we show our derived boundaries for several possible values of $\rhox$, where the red, gray, and yellow dotted curves can be compared to those derived by \cite{McDermott_2012}. For $\mx\gtrsim m_n$, the capture suppression imposed via the form factor presented in Eq.~\ref{eq:form_factor} weakens the bounds slightly, visible where the solid lines separate from the dashed in high $\mx$. Visible in the right panel on the solid blue line, the "hump" shape is the only visible effect of DM evaporation. In that region of the $\sigma$ vs $m_X$ parameter space the vast majority of the DM in the star is within the BEC near the core, yet three different effects, as detailed below, ensure that evaporation from a BEC in this case is very small, and negligible for the NS. First, since $r_{BEC}\ll R$, the particle would need to be accelerated much more than it would near the surface to escape. Further, since $\sigma$ is high in this case, $s_{evap}$ approaches 0, which further suppresses evaporation. Finally, due to the low temperature of the BEC, the required change in velocity for the DM particle to escape is further increased. An estimate of the evaporation rate from the BEC can be found in Appendix B, which in the NS case turns out to be almost completely irrelevant. Thus, the ``hump'' shape in Fig.~\ref{fig:bounds_NS} is due solely to evaporation from the relatively few MB distributed DM particles occupying states other than the BEC ground state. Note however that sensitivity is quickly restored, as increasing $\sigma$ leads to an exponential decrease of the evaporation suppression factor.

While there are some effects that may interfere with the formation of the BEC (namely Hawking Radiation, with excluded parameter space shown by square-hatching in figs. \ref{fig:bounds_NS} and \ref{fig:bounds_M1000}), evaporation from the BEC does not, as explained below. The formation of a BEC is guaranteed whenever the bosonic DM particles follow a thermal distribution and whenever $T_{DM}<T_{crit}$ (see Sec.~\ref{sec:DerivBoundsNS}). Even if DM particles evaporate from the BEC, the rest of the DM particles remain in thermal equilibrium, and the temperature of the ensemble remains unchanged. Moreover, when DM evaporation is relevant (i.e. $E\cdot t\gg 1$), the equilibration between DM capture and evaporation guarantees a constant number of DM particles in the star ($N_X=C/E$). This, in turn means that the critical temperature will not change. Therefore, DM evaporation does not spoil the conditions for the formation of a BEC, a fact we have confirmed numerically as well. The critical difference between this case and that of Hawking Radiation is that for low BH masses as would be created through this phenomenon, the temperature of the Hawking Radiation would be incredibly high, disrupting thermal equilibrium.

\begin{figure}
    \centering
    \includegraphics[width=1\linewidth]{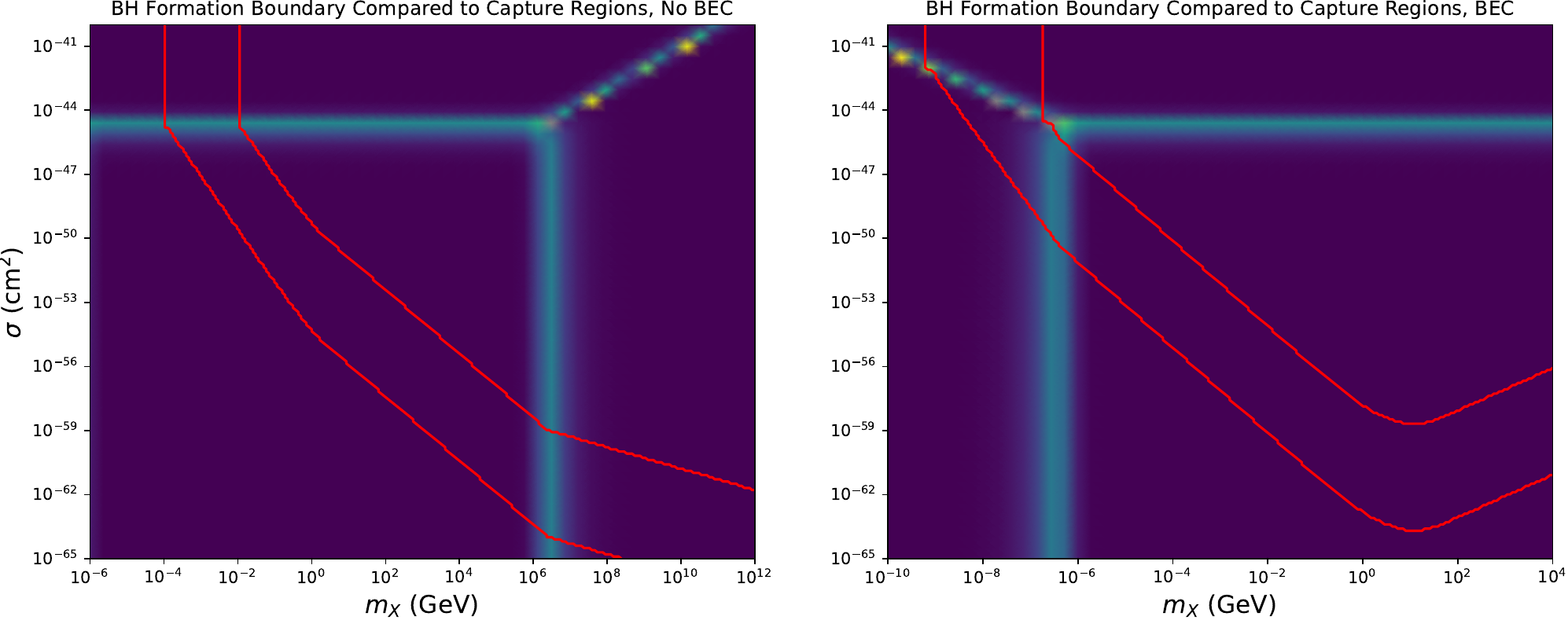}
    \caption{BH formation bounds for $\rhox=10^{13}\unit{GeV/cm^3}$ (lower red line), the upper limit of DM density we obtain using a generalized NFW profile with index $\gamma = 1.5$ (in line with \cite{immortalStars}) and $\rhox=10^{8}\unit{GeV/cm^3}$ (upper red line), the DM density approximately above which the bound enters the low-$\mx$ regions I and IV. The background is the numerically located boundary for the capture regions defined in Fig.\ref{fig:ctot_rel} - note how changes in the slope of the bound track these boundaries, except for that near $\mx=0$ which is caused by $\xi_{s}$.}
    \label{fig:bound_compare_high_rhox}
\end{figure}

The next most noticeable feature of the curves in Figure \ref{fig:bounds_NS} is undoubtedly the way they seem to change direction at similar values of $\sigma$ and $\mx$. As it turns out, this can be fully accounted for by the capture regions visible in Figure \ref{fig:ctot_rel}. For a clear visualization, in figure \ref{fig:bound_compare_high_rhox} we have plotted the $\rhox=10^{13}\unit{GeV/cm^3}$ line on the same axes as the capture regions. Notice that in the BEC case, the upper bound on $\sigma$ (red line) crosses into Regions IV and I, with an identical slope between them because their respective capture rates are equivalent (see Eqns~\ref{eq:CtotRI}-\ref{eq:CtotRIV}). Further, because of the presence of Region I, the boundary probes deeper into $\mx$ than would be possible if we assumed that capture would always saturate at $\sigma\approx 10^{-45}\unit{cm^2}$, as was done in \cite{McDermott_2012}. The effects of evaporation are neglected in Fig. \ref{fig:bound_compare_high_rhox} in favor of building a full understanding of the effects of capture regions on the shape of the boundary.

When considering values of $\rhox$ up to $\sim10^8 \unit{GeV/cm^3}$, note that the bounds are entirely within Regions IV, III, and II (see in Fig. \ref{fig:bound_compare_high_rhox} where the upper red line intersects the corner between all four regions). An interesting consequence of this is that for the values considered by \cite{McDermott_2012}, bounds calculated through single scatter capture are exact (ignoring \ref{eq:form_factor}), because Regions III and IV are the "Single Scatter Regime", where a DM particle is unlikely to interact with baryons more than once while traversing the star, so multiscatter effects are negligible (the single-scatter capture rate in \cite{McDermott_2012} reduces exactly to \ref{eq:CtotRII}-\ref{eq:CtotRIV} in the relevant regimes - only in Region I does the multiscatter formalism diverge from the single-scatter).  Further, since capture is saturated to the geometric limit in Region II, the capture rate is equal to the DM flux across the star, and the bounds lose dependence on $\sigma$. However, when $\rhox\gtrsim 10^8 \unit{GeV/cm^3}$, the bound in $\sigma - \mx$ space in the BEC case enters the left-side (low-$\mx$) Region IV as well, and more importantly,  Region I. Though the analytic capture rate in Region I is equivalent to that of Region IV, this is due to the effects of multiscatter capture, and cannot be accounted for by single scatter calculations. As such, we find that because of the placement of Region I, BEC formation allows us to probe smaller values of $\mx$, since capture does not saturate until  the boundary between Regions I and II is reached. This is visible in the right panel of Fig. \ref{fig:bounds_NS} by comparing the vertical section of the blue line to the other three - the bound is vertical when capture is saturated, and we can see that capture saturates at a higher value of $\sigma$ for the blue line compared to the other three.

Simply put, the effects of multiscatter capture and evaporation can be relevant in NS if $\rhox$ is greater than $\sim 10^8 \unit{GeV/cm^3}$, with the former allowing lower values of $\mx$ to be reached than otherwise possible, and the latter eliminating a small amount of parameter space. Second, the capture suppression effect introduced by \cite{lu2024} acts to weaken bounds when $\mx\gtrsim1\unit{GeV}$, but turns out to not be relevant when $\rhox$ is sufficiently high and a BEC forms, as DM has insufficient time to thermalize in that section of parameter space (see where the solid blue line diverges from the dotted in the right panel of Fig. \ref{fig:bounds_NS}, located entirely within the striped no-thermalization region). Finally, we find that the effects of DM evaporation are minimal (and only present for the BEC case) once one includes the evaporation suppression factor (see Eq.~\ref{eq:SuppressionFactor}), as we did throughout this work. 

\section{Application to Population III Stars}
Though NS are the most powerful class of object for this type of analysis, they have a few limitations. First, the strength of the bounds we can place depends directly on the age of the object. As the NS ages, it cools, which also strengthens the bounds we can place, but makes the object considerably more difficult to detect, especially in active areas such as the galactic core where they would be the most useful thanks to the higher DM density. So, the more compact, cooler, older, and nearer the object, the better, but the higher DM density near the object, the more luminous, the better. Undoubtedly, these requirements clash directly, and there is no perfect object that ticks all the boxes. However, instead of focusing on the former group, let us instead attempt to optimize for the latter. If we are looking for a highly luminous object in an area of high DM density, the first applicable object that springs to mind is a Population III star.

The first generation of stars, Population III (or Pop~III) stars formed from large primordial clouds of Hydrogen, allowing them to grow up to $1000$ solar masses in size\cite{Freese:2008dmdens, Barkana_2001, Bromm_2013}. In relation to this paper's analysis, these stars would be short-lived, but in return offer an incredibly large capture area due to the star's size, with a high luminosity that is expected to be detectable with JWST, especially if lensed~\citep{Zackrisson:2024PopIIIJWST}. Finally, the environment of the Pop~III star is perfect for this analysis: since these stars lived and died in the early universe, the ambient DM density was far higher, easily attaining $\rhox\sim10^{16}\unit{GeV/cm^3}$\citep[e.g.][]{Spolyar:2008dark,Freese:2008dmdens}. So, the question becomes whether the improved DM capture ability of the Population III star is enough for it to compete with a local NS in terms of constraining power, and as it turns out, the answer is complicated. For the following analysis, we use stellar parameters calculated by MESA simulations: $T_c =2\times 10^{7}$ K, $R=1.11R_\odot$, $M=1000M_\odot$.

\begin{figure}[H]
    \centering
    \includegraphics[width=1\linewidth]{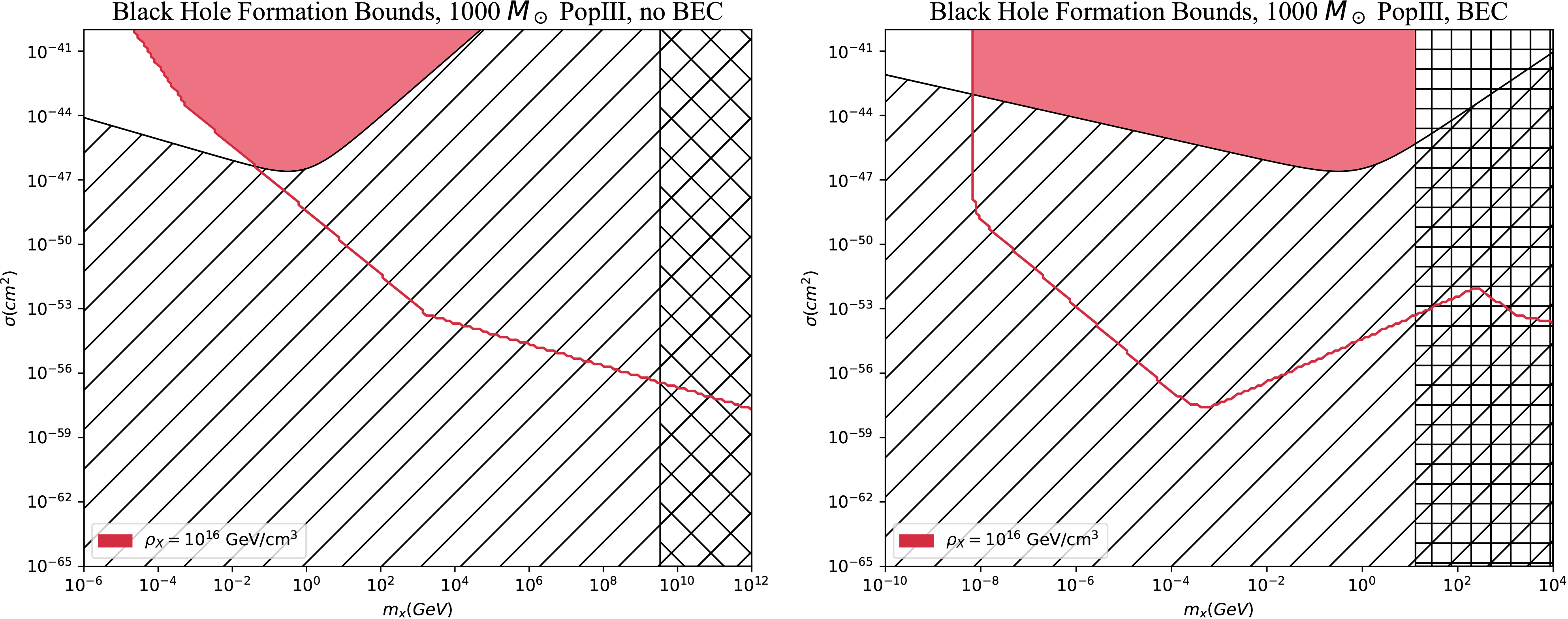}
    \caption{Bounds placed on $\sigma$ and $\mx$ by observation of a 1000 solar mass Population III star. For ease of comparison with a NS, the displayed parameter space is identical to Figure \ref{fig:bounds_NS}.}
    \label{fig:bounds_M1000}
\end{figure}

Figure \ref{fig:bounds_M1000} displays the mix of advantages and disadvantages of using a Population III star of age $10^6$ years after entering the main sequence, when it is most likely to be detected \cite{Ilie:2020PopIII}. For the case where a BEC can form (right side panel) the loss of sensitivity at $m_X\sim 10^{-8}$~GeV is due to the effects of DM evaporation, and in this case specifically to evaporation from the BEC, as this term now dominates over the total evaporation rate. This mass being so low might be surprising, as typical evaporation masses are much higher than this.

The most striking difference is the sheer amount of otherwise constrainable parameter space that is excluded by the thermalization requirement, which is unsurprising due to the short lifetime of the star. In relation to the NS another noticeable feature of the Pop III bounds is the point where the bound line becomes vertical and sensitivity is lost, but the parameter space we have selected above to match that of the NS does not tell the full story. As noted in Appendix B, if $\sigma$ becomes sufficiently high, particles that would have otherwise evaporated will tend to collide with another nucleon while exiting and remain captured. As such, if we are to consider higher values of $\sigma$ and lower $\mx$, we see that in return for a weaker ability to probe low values of $\sigma$, Population III stars are much more effective at constraining low $\mx$.

\begin{figure}
    \centering
    \includegraphics[width=1\linewidth]{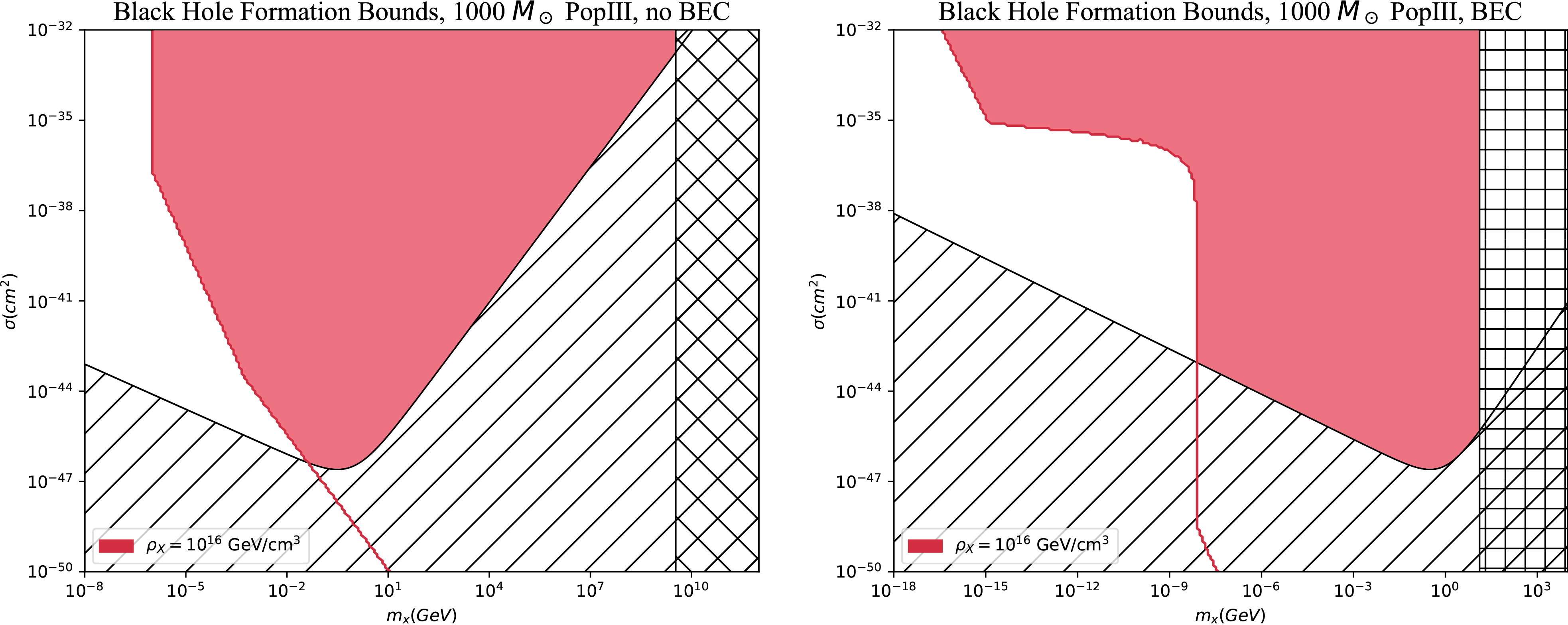}
    \caption{A second view of Figure \ref{fig:bounds_M1000}, considering higher values of $\sigma$.}
    \label{fig:M1000_highSig}
\end{figure}

In Figure \ref{fig:M1000_highSig}, the full power of the Population III star becomes clear. First, both boundaries reach low enough $\mx$, such that they traverse into region I (see Fig.~1 for an explanation of the four distinct capture regions), such that capture saturation {\it does not} limit sensitivity, even if it is above the ``naive'' geometric limit. This is most visible by noticing that the leftmost points on the red line in the right panel are not vertical. The vertical sensitivity loss in the left panel is due to $N_\chi$ becoming smaller than $N_{Cha}^{boson}$ (discussed more thoroughly below). In comparison to a NS, since locating an area of the galaxy verified to have had $\rhox=10^{13} \unit{GeV/cm^3}$ for the entire lifetime of an old NS found there is unlikely, the blue line in Figure \ref{fig:bounds_NS} is possible, but unrealistic. So, the best we could reasonably hope for is the red line, which loses constraining power at $\mx\approx 10^{-5}\unit{GeV}$. Instead, thanks to the much higher $\sigma$ at which regions III and I border region II (so capture saturates at much higher $\sigma$), mere observation of a high-mass Population III star (which would be the most easily detectable) is enough to probe as deep as $10^{-8}\unit{GeV}$ for $\sigma=10^{-40}\unit{cm^2}$, or even lower for higher $\sigma$. As alluded to previously, the vertical sensitivity loss in the left panel and at $\mx\approx10^{-8}\unit{GeV}$ in the right is not due to all the DM in the star evaporating; in fact as $\mx\rightarrow0$, $N_\chi\sim \frac{C}{E}$, meaning an equilibrium is reached. So, what that vertical line denotes is not the point at which all the DM evaporates, but the value of $\mx$ below which that equilibrium value of $N_\chi$ is insufficient to form a BH. This can occur because below $\mx\approx10^{-3}\unit{GeV}$, we enter capture region IV with rate \ref{eq:CtotRIV}, which is constant in $\mx$ when $\mx\ll m$ (to convince yourself of this, note that we fix $\rhox=n_\chi\mx$). Because $E_{BEC}\sim \mx^{-1}$ (see Eq.~\ref{eq:Ebec}) and $E_{BEC}$ dominates evaporation in this case, $N_\chi\sim \mx$, while $N^{boson}_{Cha}\sim \mx^{-2}$. So, there must be a value of $\mx$ at which $N_\chi<N^{boson}_{Cha}$, which turns out to be $\sigma-$independent and equal to approximately $10^{-8}\unit{GeV}$.

\section{Discussion and Conclusions}
In this paper, we revisited Neutron Stars as probes for Asymmetric DM, and applied a similar analysis to Population III Stars. Specifically, for the case of NSs, we improved upon the analysis done in Ref.~\cite{McDermott_2012} in several important ways. First, we consider a form factor capture suppression effect (e.g. Eq.~\ref{eq:form_factor}), and include the effects of multiscatter on DM capture. Most notably, now include in a self consistent manner Dark Matter evaporation. We do so for both DM particles in a thermal MB bath, and whenever a DM BEC condensate forms, we also include the possible effects of DM evaporation from the BEC. In the process we develop a formalism to estimate evaporation rates from the BEC (see Appenndix~B). In the NS case, we found that for  most cases considered, the effects of evaporation are irrelevant. In fact, the sensitivty loss at low masses in Fig.~\ref{fig:bounds_NS} in an area of incredibly high DM density ($\rhox=10^{13}\unit{GeV/cm^3}$), DM evaporation is relevant for a small area of parameter space, but is quickly suppressed for $\sigma\gtrsim10^{-43}\unit{cm^2}$. In Population III stars, we found that while not as effective of a low-$\sigma$ ADM detector as a NS would be given the same ambient DM density, the far higher DM density of the early universe makes Population III stars effective detectors of high-$\sigma$ low-$\mx$ DM. Further, since our analysis was performed using a Population III star that would be the easiest to detect (middle of main sequence lifetime, high-mass), we can expect to place strong bounds on ADM as soon as the first Population III stars are observed (see Figs.~\ref{fig:bounds_M1000} and Fig.~\ref{fig:M1000_highSig}). Furthermore, in line with \cite{lu2024}, we find that Neutron Stars' relativistic gravity results in suppressed capture rates when $\mx$ is not small compared to $m_n$, adding onto the disadvantages that NS have as capturers of DM. For Population III stars, these capture suppression effects are not present, resulting in the improved constraining ability we find at low-$\mx$. In particular, if a BEC forms within the Pop~III star, capture does not saturate within our considered parameter space, allowing our derived limit on $\sigma$ and $\mx$ to reach far deeper into $\mx$ than is possible with a NS.

\acknowledgments
We would like to thank the anonymous referee(s) for their insightful comments and suggestions as to how we could improve our manuscript. CI aknowledeges funding from Colgate University, via a Major Grant from Picker Interdisciplinary Sciences Institute. 

\appendix

\section{DM Particle Distribution and Evaporation Approximations}
After a DM particle is captured, traditional procedure supposes that the DM particle thermalizes with the nucleons that make up the NS, gradually losing energy and sinking deep into the star's core. When $\sigma$ is small, this is always the case, but if DM-nucleon interactions are common, then more complicated particle distributions arise. There exist two possible particle distributions that the DM may follow: when interactions are rare, DM particles reach approximate thermal equilibrium with the core of the star (hence, the "isothermal" distribution), and the DM distribution becomes "cored", with the vast majority of the DM sinking to a dense region in the center of the star. Conversely, when interactions are frequent, the DM will instead reach thermal equilibrium with the local stellar material, achieving the Local Thermal Equilibrium (or LTE) distribution. As we will confirm with \cite{Leane:2022}, \cite{Ilie:2023lbi}, and \cite{Bramante:2022}, certain values of $\sigma$ and $\mx$ can allow for a "floating" distribution, where most of the DM forms a sort of shell at the surface of the star.

Deriving the particle distribution in either case requires modelling the density and temperature profiles of the star, for which we will adopt an $n=1.5$ polytrope for the NS and $n=3$ for the Population III. The polytropic density distribution is of the form $\rho(\xi)=\rho_c\theta^n(\xi)$, where $\xi\in(0,\xi_1 )$ is a dimensionless radial variable whose maximum value is determined by $n$, and $\theta(\xi)$ is the solution to the Lane-Emden equation for the desired value of $n$:

\begin{equation}
    \frac{1}{\xi^2}\frac{d}{d\xi}\left(\xi^2 \frac{d\theta}{d\xi}\right) = -\theta^n.
    \label{eq:LaneEmden}
\end{equation}

\noindent under the initial conditions $\theta(0)=1$ and $\theta'(0)=0$. The radial variable $\xi$ is given by $\xi=\frac{r}{R}\xi_1$, where $\xi_1$ is the first positive solution to $\theta(\xi)=0$. For the $n=1.5$ polytrope, there does not exist an analytic solution to the Lane-Emden equation, but we can numerically find that $\xi_1\approx 3.654$. As such, the scale $0\leftrightarrow3.654$ becomes analogous to $0\leftrightarrow R$, since $\xi$ is to $\xi_1$ as $r$ is to $R$.

The main usefulness of the polytropic distribution is that it is simple to calculate the mass within a given radius as:

\begin{equation}
M(\xi)=4\pi\rho_c\left(\frac{\Robj}{\xi_1}\right)^3\xi^2\left(-\frac{d\theta}{d\xi}\right),
\end{equation}

\noindent which enters the gravitational potential in terms of $\xi$ as:

\begin{equation}
    \Phi(\xi) \equiv \frac{\xi_1}{R}\int_0^\xi d\xi' \frac{G M(\xi')}{\xi'^2}.
\end{equation}

\noindent For arbitrary polytrope, this simplifies to:

\begin{equation}
    \Phi(\xi) = \frac{1}{2}\frac{\xi_1}{\alpha_n}v_{esc}^2[1-\theta(\xi)],
\label{eq:PhiXi_Polytrope}
\end{equation}

\noindent where $\alpha_n \equiv \xi_1^2 \left( \frac{d\theta}{d\xi}\Bigr|_{\xi_1}\right)^2$. With \ref{eq:PhiXi_Polytrope} in hand, we are ready to approach the DM distribution within the star.

In order to determine the distribution the DM will take on, we follow \cite{Garani:2017} in introducing the parameter known as the Knudsen number, which will also permit us to interpolate between the two regimes as necessary:

\begin{equation}
    \text{Kn} = \frac{l_{\text{mfp}}}{L},
    \label{eq:KnudsenNumber}
\end{equation}

\noindent where $l_{\text{mfp}} \equiv (\sigma n_T)^{-1}$ is the approximate mean free path of DM particles within the star, $n_T$ is the average number density of neutrons in the NS, and $L$ is the length scale of the system, for which we will follow \cite{Ilie:2023lbi} in adopting $\ref{eq:rxIso}$. When $Kn\gg1$, interactions are rare and DM particles follow the LTE distribution, and conversely when $Kn\ll1$ the DM follows the Isothermal distribution. However, if $Kn\sim 1$, we will follow \cite{Garani:2017} in utilizing a transition function:

\begin{equation}
    f(\Kn) = \frac{1}{1 + \left(\Kn/0.4\right)^2},
    \label{eq:KnudsenTransitionFunction}
\end{equation}

\noindent choosing $0.4$ to numerically match expected distributions. In the intermediate regime, \cite{Garani:2017} finds the stitched DM distribution to be:

\begin{equation}
    n_X(\xi) = f(\text{Kn}) n_{X}^{\text{LTE}}(\xi) + \left[1-f(\text{Kn})\right]n_{X}^{\text{ISO}}(\xi).
    \label{eq:nx_r_Transition}
\end{equation}

\noindent where $n_{X}^{\text{LTE}}$ and $n_{X}^{\text{ISO}}$ are the distributions for either regime. Plugging this into equation \ref{eq:Evaporation_IntegralForm}, we find that we can write the full evaporation rate as the sum of those from either distribution in the same way:

\begin{equation}
    E = f(\text{Kn}) E_{\text{LTE}} + \left[1-f(\text{Kn})\right] E_{\text{ISO}},
    \label{eq:Evaporation_interpolated}
\end{equation}

\noindent 

\subsection{Isothermal Distribution:}
In their paper, \cite{Ilie:2020PopIII} show that in the isothermal limit, DM achieves the following distribution:

\begin{equation}
    n_{X}^{\text{ISO}} = n_X(0) e^{-\mx\Phi(r)/T_X}
    \label{eq:nxIso}
\end{equation}

\noindent where $n_X(0)$ is the central density of DM particles obtained by normalizing to $N_X$, and $T_X$ is the temperature of the DM particles, which is on the order of $T_c$. Defining the dimensionless variables $\mu=\frac{\mx}{m_n}$ $\Tilde{\Phi}\equiv\frac{m_n\Phi(r)}{T_c}$, and $\Theta = \frac{T_\chi}{T_c}$ in the same manner as \cite{Ilie:2020PopIII}, the DM temperature $\Theta$ is obtained by solving:

\begin{equation}
    \int_0^{\xi_1} \theta(\xi)^3 \exp{\left(-\frac{\mu}{\Theta}\Tilde{\Phi}(\xi)\right)} \left(\frac{\Theta + \mu\theta(\xi)}{\mu}\right)^\frac{1}{2} \left[\Theta - \theta(\xi)\right] \xi^2 d\xi = 0.
\end{equation}

\noindent Note that because this temperature equation is general to all objects in the isothermal limit, the definition of $\mu$ would include the proton mass in Hydrogen-composed objects rather than the neutron mass.

With \ref{eq:nxIso} in hand, we can develop an approximation for the evaporation rate from an object of arbitrary polytropic index, greatly increasing computational efficiency. Further, since integrals with large decay constants are highly imprecise to compute, the approximation allows us to compute evaporation rates in regions of parameter space at which a direct calculation of \ref{eq:Evaporation_IntegralForm} would return meaningless data. Exchanging our radial vairable for $\xi$, \ref{eq:Evaporation_IntegralForm} simplifies to:

\begin{equation}
    E_{Iso} = \frac{1}{N_X} \frac{2}{\sqrt{\pi}}\sigma \int_0^{\xi_1}dV n(\xi)u(\xi)n_X^{Iso}(\xi)e^{-\frac{v_e(\xi)^2}{v_X^2}}
\end{equation}

\noindent where $n(\xi)$ is the number density of nucleons at radius $\xi$, $u(\xi)$ is their expected velocity, and $v_X(\xi)=\sqrt{2T_X/\mx}$ is the thermal average DM velocity, which is a constant in the Isothermal case, and a function of $\xi$ in the LTE case. Assuming an $n=1.5$ polytrope, we can begin to replace the undefined functions of $\xi$ with functions of $\theta$. By definition, for arbitrary polytrope, $n(\xi)=n_c\theta^n(\xi)$, and $u(\xi) = u_c \theta^\frac{1}{2}$:

\begin{equation}
\begin{gathered}
    E_{Iso} = \frac{1}{N_X} \frac{2}{\sqrt{\pi}}\sigma n_c u_c \int_0^{\xi_1}dV s_\text{evap}(\xi)\theta^{n+\frac{1}{2}}(\xi) n_X^{Iso}(\xi)e^{-\frac{v_e(\xi)^2}{v_X^2}}, \\
    E_{Iso} = \frac{n_X(0)}{N_X} \frac{2}{\sqrt{\pi}}\sigma n_c u_c \int_0^{\xi_1}dV s_\text{evap}(\xi) \theta^{n+\frac{1}{2}}(\xi) e^{\frac{-\mx\Phi(\xi)}{T_X}-\frac{v_e(\xi)^2}{v_X^2}}, 
\end{gathered}
\end{equation}

\noindent Focusing on the exponential term, we can take advantage of the polytropic definition of $\Phi(\xi)$ to find $v_e(\xi)=\vesc^2(1+\frac{\xi_1}{2}\theta(\xi))$:

\begin{equation}
\begin{gathered}
    \frac{-\mx\Phi(\xi)}{T_X}-\frac{v_e(\xi)^2}{v_X^2},\\
    -\frac{\mx}{T_X}\frac{1}{2}\frac{\xi_1}{\alpha_n}\vesc^2(1-\theta(\xi))-\frac{\vesc^2}{v_X^2}\left(1+\frac{\xi_1}{2}\theta(\xi)\right),\\
    -\frac{\vesc^2}{v_X^2}\left(\frac{\xi_1}{\alpha_n}(1-\theta(\xi))+1+\frac{\xi_1}{2}\theta(\xi)\right),\\
    -\frac{\vesc^2\xi_1}{v_X^2}\left( \frac{1}{\alpha_n} + \frac{1}{\xi_1}+ \theta(\xi)\left[\frac{1}{2}-\frac{1}{\alpha_n}\right] \right).
\end{gathered}
\end{equation}

\noindent Letting $\Xi(\xi) \equiv \int_0^{\xi_1}dV s_\text{evap}(\xi)\theta^{n+\frac{1}{2}}\exp{\left( -\frac{\vesc^2\xi_1}{v_X^2}\left( \frac{1}{\alpha_n} + \frac{1}{\xi_1} \theta(\xi)\left[\frac{1}{2}-\frac{1}{\alpha_n}\right] \right)\right)}$, the full equation for $E_{Iso}$ becomes:

\begin{equation}
    E_{Iso} = \frac{n_X(0)}{N_X} \frac{2}{\sqrt{\pi}}\sigma n_c u_c \Xi(\xi),
    \label{eq:Evap_Iso}
\end{equation}

\noindent where we note that in the case of $n=3$, $\alpha_n\approx 2$, $\Xi$ becomes $\xi$-independent, and equation C8 from \cite{Ilie:2020PopIII} is recovered.

\subsection{LTE Distribution}
In the Local Thermal Equilibrium regime, \cite{1990ApJ...352..654G} find that the DM distribution becomes:

\begin{equation}
    n_X^{LTE}(r) = n_{X}(0)\left(\frac{T(r)}{T_c}\right)^{3/2}\exp\left(-\int_0^r\frac{\alpha(r')\frac{dT}{dr'} + m_\chi \frac{d\Phi}{dr'}}{T(r')} dr'\right),
\label{eq:nxLTE_general}
\end{equation}

As derived in \cite{Ilie:2023lbi}, this simplifies to the following when we assume an $n=1.5$ polytrope and $T(r)\sim \theta$, as is the case for NS:

\begin{equation}
    n_X^{LTE}(r)= n_X(0) \theta^{p}(\xi), \quad p = \frac{\xi_1}{\alpha_n}\frac{E_e}{T_c} + \frac{3}{2} - \alpha_0
    \label{eq:nxLTE}
\end{equation}

\noindent where $E_e=\frac{1}{2}\mx \vesc^2$ is the escape energy from the surface of the star. As it turns out, if only one species of nucleon is considered, $\alpha(r)$ becomes a constant $\alpha_0$, which can be approximated as shown in Appendix A of \cite{1990ApJ...352..654G}.

Similarly to the Isothermal case, by simplifying the integral, we can increase the accuracy of our bounds (as integrating functions with high decay constants is imprecise) by deriving an approximation for the evaporation coefficient using \ref{eq:nxLTE}. Exchanging our radial variable for $\xi$, the triple integral in \ref{eq:Evaporation_IntegralForm} simplifies substantially to:

\begin{equation}
    E_{LTE} = \frac{1}{N_X} \frac{2}{\sqrt{\pi}}\sigma \int_0^{\xi_1}dV n(\xi)u(\xi)n_X^{LTE}(\xi)e^{-\frac{v_e(\xi)^2}{v_X(\xi)^2}}
\end{equation}

\noindent where definitions of all variables expect those labeled with LTE are identical to the Isothermal case. Plugging in \ref{eq:nxLTE} and other known functions in the same manner as was done for the Isothermal case, we can carry out the following sequence of simplifications:

\begin{equation}
\begin{gathered}
    E_{LTE} = \frac{1}{N_X} \frac{2}{\sqrt{\pi}}\sigma n_c u_c \int_0^{\xi_1}dV s_\text{evap}(\xi)\theta^{n+\frac{1}{2}}(\xi) n_X^{LTE}(\xi)e^{-\frac{v_e(\xi)^2}{v_X(\xi)^2}}, \\
    = \frac{n_X(0)}{N_X} \frac{2}{\sqrt{\pi}}\sigma n_c u_c  \int_0^{\xi_1}dV s_\text{evap}(\xi)\theta^{n+\frac{1}{2}} \exp{\left(\ln{p} - \frac{v_e(\xi)^2}{v_X(\xi)^2} \right)}\\
    =  \frac{n_X(0)}{N_X} \frac{2}{\sqrt{\pi}}\sigma n_c u_c \int_0^{\xi_1}dV s_\text{evap}(\xi)\theta^{n+\frac{1}{2}}\exp{\left(\left[ p - \frac{v_e(\xi)^2}{v_X(\xi)^2\ln\theta} \right] \ln\theta\right)}\\
    = \frac{n_X(0)}{N_X} \frac{2}{\sqrt{\pi}}\sigma n_c u_c \int_0^{\xi_1}dV s_\text{evap}(\xi)\theta^{p_{\text{ev}}}, \quad p_\text{ev} = p + n + \frac{1}{2} - \frac{v_e(\xi)^2}{v_X(\xi)^2\ln\theta}
\end{gathered}
\label{eq:Evap_LTE}
\end{equation}

\noindent where we have assumed that the star has a uniform central DM density $n_X(0)$ that both the LTE and Isothermal distributions are normalized to. Of course, $N_X$ cannot be calculated without knowing the evaporation coefficient, which we work around by observing that $\int_0^{\xi_1}dV n_X(\xi) = N_X$.

\section{Evaporation from a Bose-Einstein Condensate}
In this section, we will approximate the evaporation rate from a Bose-Einstein Condensate of DM particles, which eventually is included in all of our numeric calculations of the total evaporation rates (see Eq.~\ref{eq:Evap_sum}). We estimate the evaporation rate from a BEC from its integral form presented in Eq.~\ref{eq:Evaporation_IntegralForm}. The BEC is modeled as a small sphere of iso-velocity particles. The per-particle energy of the ground state confined to a box of side length L is given by:

\begin{equation}
    \epsilon_0 = \frac{h^2}{8m_\chi L^2}
\end{equation}

A good order of magnitude estimate for this energy in our case can be attained by letting $r_{BEC}=L$. So, assuming all of this energy is kinetic, the velocity of a particle in the ground state of the BEC is given by:

\begin{equation}
    v_{BEC}=\sqrt{\frac{3h^2}{4m_\chi^2 r_{BEC}^2}}.
\end{equation}

Since every particle in the BEC has roughly this velocity, the calculation of $E$ becomes much simpler. Further, because this sphere of particles is very small in relation to the star's radius (valid to less than 1\% in all considered parameter space), we will simplify our model even further as to consider that all DM particles are located at the center of the star for the purposes of BEC evaporation. So, $T(r)=T_c$, $n(r)=n_c$, $f_\chi(\textbf{w})=\delta^3(w-v_{BEC})$, and $n_\chi(r)=\delta^3(r)N_\chi$, where $\delta^3$ is the 3 dimensional Dirac delta function. Under these conditions, \ref{eq:Evaporation_IntegralForm} simplifies greatly to:

\begin{equation}
    E_{BEC} = s_{evap}(0)\int_{v_e(0)}^\infty \mathcal{R}^+ (v_{BEC}\rightarrow v)~dv \equiv s_{evap}(0)\Omega^+(v_{BEC}),
\end{equation}

\noindent in which the full form of $\Omega^+$ is given by \cite{1987ApJ...321..560G}. So, to close approximation, the BEC evaporation is defined fully by the interplay between the suppression factor and the chance of upscattering past the escape velocity. This interplay is critically important in the low-$\mx$ high-$\sigma$ region, as $\Omega^+\rightarrow\infty$ as $\mx\rightarrow0$, but $s_{evap}\rightarrow0$ for large $\sigma$. In order to determine which term is dominant in this region, we will examine the behavior of both terms in their relevant limits. 

In its entirety, the form of $\Omega^+$ is rather complicated, but for both the NS and the Population III star cases, the escape velocity from the stars' core is very large, so evaporation tends to occur only for very small $\mu$. In this limit, with $m$ as the nucleon mass and $v_e$ as the escape velocity from the center of the star, $\Omega^+$ reduces to:

\begin{equation}
    \Omega^+(v_{BEC})\approx \frac{n\sigma(\sqrt{\pi}-2)}{2\pi\sqrt{\frac{2m}{T_c}}v_{BEC}\mu} e^{-\frac{m(v_e+v_{BEC})^2}{8T_c}}\left(v_e-v_{BEC}-e^{\frac{m v_e v_{BEC}}{2T_c}} \right).
    \label{eq:Ebec}
\end{equation}

\noindent Note the scaling with $\mu^{-1}$. However, this expression is only valid in the $\mu\ll1$ limit, so when this is not the case, we use the full form of $\Omega^+$ as given by \cite{1987ApJ...321..560G}. Conversely, the leading order term of $s_{evap}$ as the optical depth from the star's center $\tau_0\equiv2R\sigma \bar n$ (where $\bar n$ is the average number density of nucleons in the whole star) is:

\begin{equation}
    s_{evap}\approx e^{-\tau_0}\tau_0^{-\frac{\hat\phi}{3}}\text{O}[\tau_0^{-\frac{5}{4}}]
    \label{sevap_approx}
\end{equation}

While $\Omega^+$ decays as $\mx^{-1}$, $s_{evap}$ decays exponentially with $\sigma$. The competition between those two terms will dictate when, if ever, evaporation from a BEC becomes relevant. We note that at sufficiently high $\sigma$ the suppression factor will take over, and $E_{BEC}\simeq s_{evap}\Omega^+(v_{BEC})\approx0$. 
 We have numerically determined this to occur at approximately $\sigma=10^{-44}\text{ cm}^2$ for the NS (visible in the right panel of Fig \ref{fig:bounds_NS} as the horizontal section of the $\rhox=10^{13} \text{ GeV/cm}^3$ line), and $\sigma=10^{-35}\text{ cm}^2$ for the Population III. As alluded to before, in all of our numerical work we now include the possible effects of evaporation from a BEC, estimated as described in this appendix.

\bibliographystyle{JHEP}
\bibliography{refsDM}

\providecommand{\href}[2]{#2}\begingroup\raggedright\begin{thebibliography}{10}

\bibitem{1933AcHPh...6..110Z}
F.~{Zwicky}, \emph{{Die Rotverschiebung von extragalaktischen Nebeln}}, {\emph{Helvetica Physica Acta} {\bfseries 6} (1933) 110}.

\bibitem{10.1093/mnras/stad3393}
I.~Banik, C.~Pittordis, W.~Sutherland, B.~Famaey, R.~Ibata, S.~Mieske et~al., \emph{{Strong constraints on the gravitational law from Gaia DR3 wide binaries}}, \href{https://doi.org/10.1093/mnras/stad3393}{\emph{Monthly Notices of the Royal Astronomical Society} {\bfseries 527} (2023) 4573} [\href{https://arxiv.org/abs/https://academic.oup.com/mnras/article-pdf/527/3/4573/53896686/stad3393.pdf}{{\ttfamily https://academic.oup.com/mnras/article-pdf/527/3/4573/53896686/stad3393.pdf}}].

\bibitem{Kaplan_2009}
D.~E. Kaplan, M.~A. Luty and K.~M. Zurek, \emph{Asymmetric dark matter}, \href{https://doi.org/10.1103/physrevd.79.115016}{\emph{Physical Review D} {\bfseries 79} (2009) }.

\bibitem{Aprile_2017}
E.~Aprile, J.~Aalbers, F.~Agostini, M.~Alfonsi, F.~D. Amaro, M.~Anthony et~al., \emph{The xenon1t dark matter experiment}, \href{https://doi.org/10.1140/epjc/s10052-017-5326-3}{\emph{The European Physical Journal C} {\bfseries 77} (2017) }.

\bibitem{pettus2015dmicecurrentstatusfuture}
W.~C. Pettus, \emph{Dm-ice: Current status and future prospects},  2015.

\bibitem{kluck2017searchlowmassdarkmatter}
H.~Kluck, G.~Angloher, P.~Bauer, A.~Bento, C.~Bucci, L.~Canonica et~al., \emph{Search for low-mass dark matter with the cresst experiment},  2017.

\bibitem{Ilie:2023lbi}
C.~Ilie, C.~Levy and J.~Diks, \emph{{The effectiveness of exoplanets and Brown Dwarfs as sub-GeV Dark Matter detectors}}, \href{https://doi.org/10.1088/1475-7516/2024/04/082}{\emph{JCAP} {\bfseries 04} (2024) 082} [\href{https://arxiv.org/abs/2312.13979}{{\ttfamily 2312.13979}}].

\bibitem{Leane:2020wob}
R.~K. Leane and J.~Smirnov, \emph{Exoplanets as sub-gev dark matter detectors}, \href{https://doi.org/10.1103/PhysRevLett.126.161101}{\emph{Phys. Rev. Lett.} {\bfseries 126} (2021) 161101}.

\bibitem{Ilie:2020PopIII}
C.~Ilie, C.~Levy, J.~Pilawa and S.~Zhang, \emph{Constraining dark matter properties with the first generation of stars}, \href{https://doi.org/10.1103/PhysRevD.104.123031}{\emph{Phys. Rev. D} {\bfseries 104} (2021) 123031}.

\bibitem{Freese:2008cap}
K.~Freese, D.~Spolyar and A.~Aguirre, \emph{{Dark Matter Capture in the first star: a Power source and a limit on Stellar Mass}}, \href{https://doi.org/10.1088/1475-7516/2008/11/014}{\emph{JCAP} {\bfseries 0811} (2008) 014} [\href{https://arxiv.org/abs/0802.1724}{{\ttfamily 0802.1724}}].

\bibitem{Horowitz:2020axx}
C.~J. Horowitz, \emph{{Nuclear and dark matter heating in massive white dwarf stars}}, \href{https://doi.org/10.1103/PhysRevD.102.083031}{\emph{Phys. Rev. D} {\bfseries 102} (2020) 083031} [\href{https://arxiv.org/abs/2008.03291}{{\ttfamily 2008.03291}}].

\bibitem{Dasgupta:2020dik}
B.~Dasgupta, A.~Gupta and A.~Ray, \emph{{Dark matter capture in celestial objects: light mediators, self-interactions, and complementarity with direct detection}}, {\emph{arXiv e-prints} (2020) } [\href{https://arxiv.org/abs/2006.10773}{{\ttfamily 2006.10773}}].

\bibitem{Bramante:2017}
J.~{Bramante}, A.~{Delgado} and A.~{Martin}, \emph{{Multiscatter stellar capture of dark matter}}, \href{https://doi.org/10.1103/PhysRevD.96.063002}{\emph{\prd} {\bfseries 96} (2017) 063002} [\href{https://arxiv.org/abs/1703.04043}{{\ttfamily 1703.04043}}].

\bibitem{Ilie:2020Comment}
C.~Ilie, J.~Pilawa and S.~Zhang, \emph{Comment on ``multiscatter stellar capture of dark matter''}, \href{https://doi.org/10.1103/PhysRevD.102.048301}{\emph{Phys. Rev. D} {\bfseries 102} (2020) 048301}.

\bibitem{Nguyen:2022NS}
T.~T.~Q. Nguyen and T.~M.~P. Tait, \emph{{Bounds on long-lived dark matter mediators from neutron stars}}, \href{https://doi.org/10.1103/PhysRevD.107.115016}{\emph{Phys. Rev. D} {\bfseries 107} (2023) 115016} [\href{https://arxiv.org/abs/2212.12547}{{\ttfamily 2212.12547}}].

\bibitem{Baryakhtar:2017}
M.~{Baryakhtar}, J.~{Bramante}, S.~W. {Li}, T.~{Linden} and N.~{Raj}, \emph{{Dark Kinetic Heating of Neutron Stars and an Infrared Window on WIMPs, SIMPs, and Pure Higgsinos}}, \href{https://doi.org/10.1103/PhysRevLett.119.131801}{\emph{\prl} {\bfseries 119} (2017) 131801} [\href{https://arxiv.org/abs/1704.01577}{{\ttfamily 1704.01577}}].

\bibitem{Bell:2020}
N.~F. {Bell}, G.~{Busoni}, S.~{Robles} and M.~{Virgato}, \emph{{Improved Treatment of Dark Matter Capture in Neutron Stars}}, {\emph{arXiv e-prints} (2020) arXiv:2004.14888} [\href{https://arxiv.org/abs/2004.14888}{{\ttfamily 2004.14888}}].

\bibitem{Bell:2018pkk}
N.~F. Bell, G.~Busoni and S.~Robles, \emph{{Heating up Neutron Stars with Inelastic Dark Matter}}, \href{https://doi.org/10.1088/1475-7516/2018/09/018}{\emph{JCAP} {\bfseries 09} (2018) 018} [\href{https://arxiv.org/abs/1807.02840}{{\ttfamily 1807.02840}}].

\bibitem{McDermott_2012}
S.~D. McDermott, H.-B. Yu and K.~M. Zurek, \emph{Constraints on scalar asymmetric dark matter from black hole formation in neutron stars}, \href{https://doi.org/10.1103/physrevd.85.023519}{\emph{Physical Review D} {\bfseries 85} (2012) }.

\bibitem{lu2024}
C.-T. Lu, A.~K. Mishra and L.~Wu, \emph{Constraining bosonic dark matter-baryon interactions from neutron star collapse},  2024.

\bibitem{1988ApJ...328..919G}
A.~{Gould}, \emph{{Direct and Indirect Capture of Weakly Interacting Massive Particles by the Earth}}, \href{https://doi.org/10.1086/166347}{\emph{\apj} {\bfseries 328} (1988) 919}.

\bibitem{1987ApJ...321..560G}
A.~{Gould}, \emph{{Weakly Interacting Massive Particle Distribution in and Evaporation from the Sun}}, \href{https://doi.org/10.1086/165652}{\emph{\apj} {\bfseries 321} (1987) 560}.

\bibitem{2024IlieCapture}
C.~{Ilie}, \emph{{Closed-form Expressions for Multiscatter Dark Matter Capture Rates}}, \href{https://doi.org/10.3847/1538-4357/ad5556}{\emph{\apj} {\bfseries 970} (2024) 159} [\href{https://arxiv.org/abs/2402.07713}{{\ttfamily 2402.07713}}].

\bibitem{Garani_2022}
R.~Garani and S.~Palomares-Ruiz, \emph{Evaporation of dark matter from celestial bodies}, \href{https://doi.org/10.1088/1475-7516/2022/05/042}{\emph{Journal of Cosmology and Astroparticle Physics} {\bfseries 2022} (2022) 042}.

\bibitem{Garani:2017}
R.~Garani and S.~Palomares-Ruiz, \emph{Dark matter in the sun: scattering off electrons vs nucleons}, \href{https://doi.org/10.1088/1475-7516/2017/05/007}{\emph{Journal of Cosmology and Astroparticle Physics} {\bfseries 2017} (2017) 007}.

\bibitem{1985Spergel}
D.~N. {Spergel} and W.~H. {Press}, \emph{{Effect of hypothetical, weakly interacting, massive particles on energy transport in the solar interior}}, \href{https://doi.org/10.1086/163336}{\emph{\apj} {\bfseries 294} (1985) 663}.

\bibitem{Navarro_1996}
J.~F. Navarro, C.~S. Frenk and S.~D.~M. White, \emph{The structure of cold dark matter halos}, \href{https://doi.org/10.1086/177173}{\emph{The Astrophysical Journal} {\bfseries 462} (1996) 563}.

\bibitem{immortalStars}
I.~John, R.~K. Leane and T.~Linden, \emph{Dark branches of immortal stars at the galactic center},  2024.

\bibitem{Freese:2008dmdens}
K.~Freese, P.~Gondolo, J.~Sellwood and D.~Spolyar, \emph{{Dark Matter Densities during the Formation of the First Stars and in Dark Stars}}, \href{https://doi.org/10.1088/0004-637X/693/2/1563}{\emph{Astrophys. J.} {\bfseries 693} (2009) 1563} [\href{https://arxiv.org/abs/0805.3540}{{\ttfamily 0805.3540}}].

\bibitem{Barkana_2001}
R.~Barkana and A.~Loeb, \emph{In the beginning: the first sources of light and the reionization of the universe}, \href{https://doi.org/10.1016/s0370-1573(01)00019-9}{\emph{Physics Reports} {\bfseries 349} (2001) 125–238}.

\bibitem{Bromm_2013}
V.~Bromm, \emph{Formation of the first stars}, \href{https://doi.org/10.1088/0034-4885/76/11/112901}{\emph{Reports on Progress in Physics} {\bfseries 76} (2013) 112901}.

\bibitem{Zackrisson:2024PopIIIJWST}
E.~{Zackrisson}, A.~{Hultquist}, A.~{Kordt}, J.~M. {Diego}, A.~{Nabizadeh}, A.~{Vikaeus} et~al., \emph{{The detection and characterization of highly magnified stars with JWST: prospects of finding Population III}}, \href{https://doi.org/10.1093/mnras/stae1881}{\emph{\mnras} {\bfseries 533} (2024) 2727} [\href{https://arxiv.org/abs/2312.09289}{{\ttfamily 2312.09289}}].

\bibitem{Spolyar:2008dark}
D.~Spolyar, K.~Freese and P.~Gondolo, \emph{{Dark matter and the first stars: a new phase of stellar evolution}}, \href{https://doi.org/10.1103/PhysRevLett.100.051101}{\emph{Phys. Rev. Lett.} {\bfseries 100} (2008) 051101} [\href{https://arxiv.org/abs/0705.0521}{{\ttfamily 0705.0521}}].

\bibitem{Leane:2022}
R.~K. Leane and J.~Smirnov, \emph{Floating dark matter in celestial bodies}, .

\bibitem{Bramante:2022}
J.~Bramante, J.~Kumar, G.~Mohlabeng, N.~Raj and N.~Song, \emph{Light dark matter accumulating in terrestrial planets: Nuclear scattering}, .

\bibitem{1990ApJ...352..654G}
A.~{Gould} and G.~{Raffelt}, \emph{{Thermal Conduction by Massive Particles}}, \href{https://doi.org/10.1086/168568}{\emph{\apj} {\bfseries 352} (1990) 654}.

\end{thebibliography}\endgroup

\end{document}